# A probe into the chaotic nature of total ozone time series by Correlation Dimension method

Goutami Bandyopadhyay and Surajit Chattopadhyay**

1/19 Dover Place

Kolkata-700 019

West Bengal

India

**Corresponding author: e-mail surajit_2008@yahoo.co.in

## Abstract

Present paper deals with mean monthly total ozone concentration over Arosa, Switzerland. The basic rationale is to investigate the existence of chaos within the relevant time series. Method of correlation dimension is adopted here as the research methodology. After a rigorous investigation a low dimensional chaos is identified within the time series pertaining to mean monthly total ozone concentration over Arosa, Switzerland.

## 1. Introduction

Management of weather and environment –sensitive endeavors very frequently requires making decisions in the face of uncertainty. In most of the cases, the decision making process is dynamic in nature [19]. This implies that the future decisions depend heavily upon the decisions already taken. In all such cases, a proper analysis of the relevant time series is an absolute necessity. Traditionally, statistical approaches are adopted to analyze the time series pertaining to weather and environment related phenomena and are mostly based upon several tacit assumptions [18, 19]. But, such assumptions sometimes lead to erroneous conclusions if the underlying processes are non-linear, complex and highly dynamic in nature.

To analyze time series data in terms of non-linear dynamics, chaos theory plays the role of most direct link [8]. Weigend and Gershenfeld (1993) [Ref 17] discussed Impact of past decisions upon the future decisions in the situation of intrinsic chaos. Many areas of science, including biology, physiology, and medicine; geo- and astrophysics, hydrology, as well as the social sciences and finance have been diagnosed with chaotic properties [e.g. 2, 4, 5, 9, 10, 11, 15 and many others]. In recent decades, research on non-linear deterministic dynamics has created new insights in the problems associated with complex phenomena [12]. In non-linear dynamics, characterization of chaos from real world observations is a difficult problem [4]. Systematic study of chaos started in the 1960s [13, 14]. It started because the linear techniques dominating the field of applied mathematics found inadequate while dealing with chaotic phenomena; and in the case of amazing irregularities within non-linear deterministic systems, the linear methods identified them as stochastic.

Study of potentially chaotic behavior can be divided into three groups: identification of chaotic behavior, modeling and prediction, and control [3]. Present study involves the first area. In recent years, identification of chaotic behavior of the geophysical processes has gained considerable attention [e.g. 1, 10, 11, 15]. But, in most of the cases, the studies involved hydro-geological processes or climatic processes.

Present study is concerned about mean monthly total ozone time series over Arosa, Switzerland. The photochemical processes leading to formation of ozone (O3) are highly complex in nature. Ozone is a secondary pollutant and is not usually emitted directly from stacks, but instead is formed in the atmosphere as a result of reactions between other pollutants emitted mostly by industries and automobiles. The ozone precursors are generally divided into two groups, namely oxides of nitrogen (NOX) and volatile organic components (VOC) like evaporative solvents and other hydrocarbons. In suitable ambient meteorological condition (e.g. warm, sunny/clear day) ultraviolet radiation (UV) causes the precursors to interact photochemically in a set of reactions that result in the formation of ozone. Ozone absorbs both incoming solar radiation in the UV and visible region, and terrestrially emitted infrared (IR) radiation. Stratospheric ozone absorbs about 12Wm-2 of solar radiation and 8Wm-2 of terrestrial IR radiation. Almost 60% of this absorbed IR is radiated. Because of its capability to absorb the incoming radiation, the stratospheric ozone is a major source of stratospheric heating, which further heats the troposphere. Again, because of radiation of IR the tropopause gets some cooling. Thus, stratospheric ozone exerts both heating and cooling effect on the land-troposphere system. Total ozone includes both tropospheric and stratospheric ozone. Since the stratospheric ozone varies on different time scales, the influence on surface UV-B and related tropospheric ozone chemistry also acts on different time scales. Thus, the total ozone time series is characterized by huge non-linearity attributed to meteorological variables, tropospheric ozone, and stratospheric ozone. Present contribution aims to adopt correlation dimension method [4, 5, 9] to detect the intrinsic chaos within the time series pertaining to the mean monthly total ozone time series over Arosa, Switzerland.

## 2 Methodology

Plethoras of literature are available where the theoretical concepts underlying the methodologies for the detection and modeling of nonlinear dynamical and chaotic components have been discussed [e.g. 4, 6, 7, 16]. Present paper adopts the method of correlation dimension method [9] to detect the presence of chaos in the time series pertaining to the mean monthly total ozone data over Arosa, Switzerland between 1932 and 1971. The data are collected from http://www.robhyndman.info/TSDL/monthly/arosa.dat. The measurements are taken in Dobson Units (DU) (300 DU=1layer of 3mm if the whole ozone column is taken at the sea level with standard conditions).

### 2.1 Correlation dimension method

According to Sivakumar (2001) [Ref 9] the correlation dimension is a representation of the variability or irregularity of a process and furnishes information on the number of dominant variables present in the evolution of the corresponding dynamical system. Computation of correlation dimension is also known as Grassberger-Procassia (GP) method. In this method a scalar time series with N entries is first considered as

$$\{X(t): t = 1, 2, 3, \ldots, N-1, N\}$$

The phase space is reconstructed as

$$Y_j = \{X_j, X_{j+\tau}, X_{j+2\tau}, \ldots\ldots, X_{j+(m-1)\tau}\}$$

Where,

$$j = 1, 2, \ldots\ldots, \frac{N-(m-1)\tau}{\Delta t}$$

Where, τ is the delay time, **m** is called the embedding dimension

For an m-dimensional phase space the correlation dimension is computed in terms of

$$C(l) = \frac{1}{N(N-1)} \sum_{i,j=1; i \neq j}^{N} \theta\left(l - |\underline{x}_i - \underline{x}_j|\right) \qquad \ldots \ldots (1)$$

Where, $\theta(x)$ is the Heaviside step function defined as:

$$\theta(p) = 0 \text{ for } p \leq 0 \text{ and } \theta(p) = 1 \text{ for } p > 0$$

Where, $p = l - |\underline{x}_i - \underline{x}_j|$

The quantity $\lim_{N \to \propto} C(l)$ is termed as correlation integral.

The scaling distance $l$ gives the estimate $\gamma$ of the attractor dimension, $C(l) \propto l^{\gamma}$. The quantity $\gamma$ is termed as correlation exponent and the saturation value of the correlation exponent is called the correlation dimension of the attractor. When the embedding dimension is unknown, $C(l)$ is computed for a number of values of **m**. If the slope of $\log C(l)$ versus $\log(l)$ converges to $\gamma$, the system is identified as a chaotic system. The slope is generally computed using the method of least squares fit of a straight line over a certain range of $l$ called the scaling region.

Phase space reconstruction as mentioned earlier requires a time delay coordinate $\tau$. The most popular choice of time delay coordinate is the time at which the autocorrelation function has its first zero [16].

## 2.2 Implementation procedure

The dataset explored in the present study consists of 40 years (1932-1971), that is, 480 months. Thus, the scalar time series of mean monthly total ozone concentration over Arosa contains 480 data points. The variation in the mean monthly total ozone data is presented in Fig.01and some important statistics regarding the data are presented in Table01. Now the correlation functions and the correlation exponents are computed for the whole data series. Fig.02 shows the correlation function up to several lags.

The delay time, $\tau$**,** for the phase space reconstruction is computed using the auto-correlation function method and is taken as the lag time at which the auto-correlation function first crosses the zero line [9]. Fig.02 shows the autocorrelation function for the mean monthly total ozone time series over Arosa, Switzerland and is found to cross the 0-line at time lag 4. Thus, 4 is taken as the delay time in the phase-space reconstruction process. In the next step, the $\log C(l)$ values are plotted against $\log (l)$. Values of embedding dimension are considered up to 19. The $\log C(l)$ versus $\log (l)$ plots are presented in Figs 03 (a) and (b). In Fig.03 (a), embedding dimensions up to 9 are presented and embedding dimensions from 10 to 19 are plotted in Fig.03 (b).

The $\log C(l)$ versus $\log(l)$ plots indicate clear scaling regions that allow reasonably precise estimates of the correlation exponents. Fig.04 shows the relationship between correlation exponent values and the embedding dimension values for the total ozone time series. Fig.04 shows that the correlation exponents saturate at embedding dimension 17 and the saturation value of the correlation dimension is 1. Thus, it can be concluded that correlation dimension of the attractor is 1. Since the correlation dimension is finite and low it can be said that the time series of the mean monthly total ozone concentration over Arosa, Switzerland is characterized by a lo-dimensional chaos.

Since the correlation dimension value is very low, it can be inferred that though the system is chaotic, the degree of variability is not vary high. Thus, complexity is there within the system but the degree of complexity is not very high. This is further supported by the low value of coefficient of variation (=0.11) (Table 01). The correlation dimension value 1 indicates that the number of dominant variables involved in the dynamics of the mean monthly total ozone time series is 1.

## 3 Conclusion

Present study identifies the existence of low dimensional chaos within the time series of mean monthly total ozone concentration over Arosa, Switzerland. In future, it can be studied how the chaotic behavior changes with the variability in the temporal scale.

## Acknowledgement

Authors sincerely acknowledge the inspiration and assistance in various ways from Mr. Sujit Chattopadhyay and Mrs Kamala Chattopadhyay while carrying out the research and preparing the manuscript.

**Table 01**. Tabular presentation of the data statistics

| Parameter | Measure |
|---|---:|
| Data points | 480 |
| Mean | 337.77083 |
| Standard deviation | 37.992604 |
| Coefficient of variation | 0.1124 |
| Maximum value | 430 |
| Minimum value | 266 |

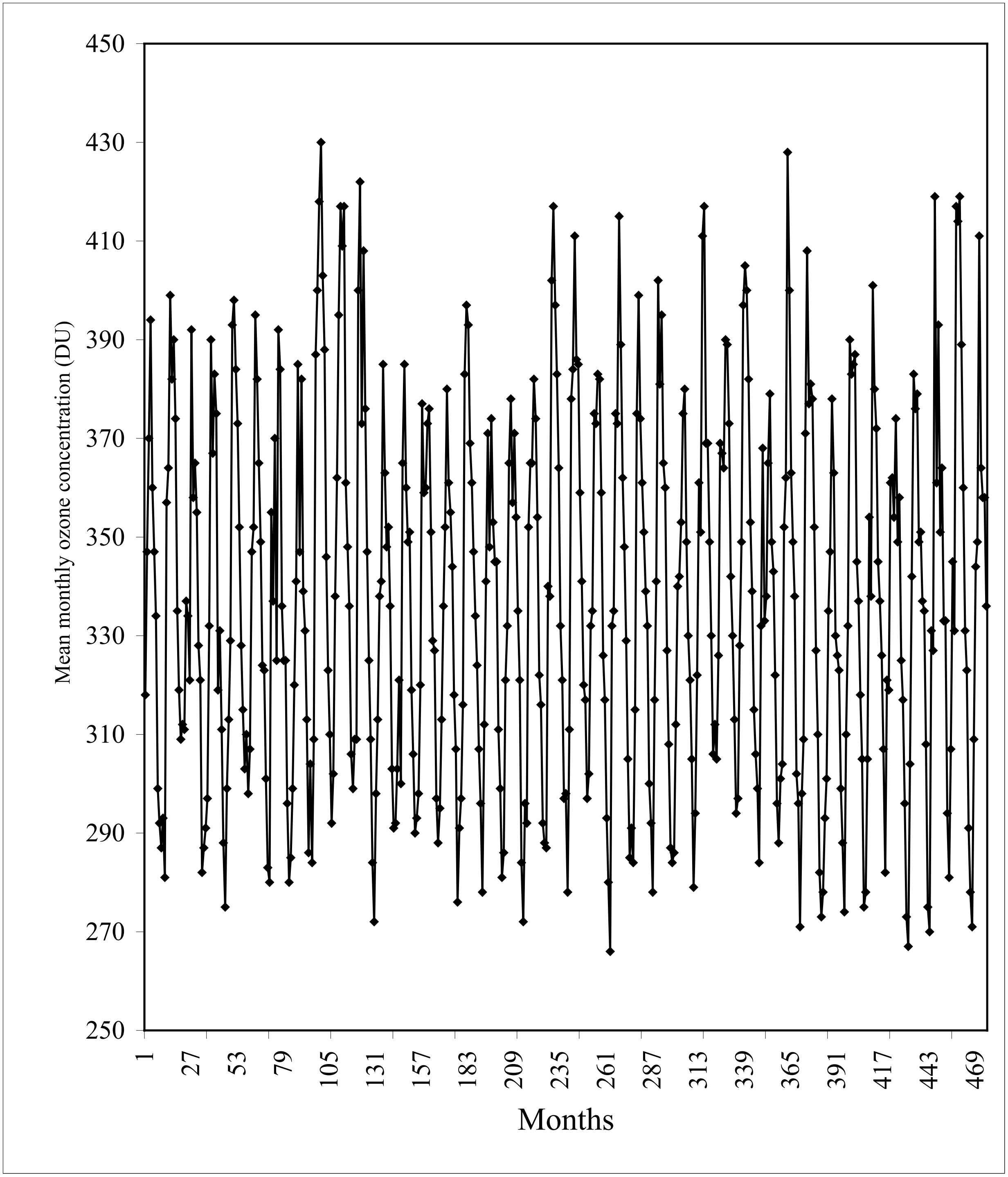

**Fig.01**- Mean monthly ozone concentration in Dobson Unit over Arosa, Switzerland during 1932-1971

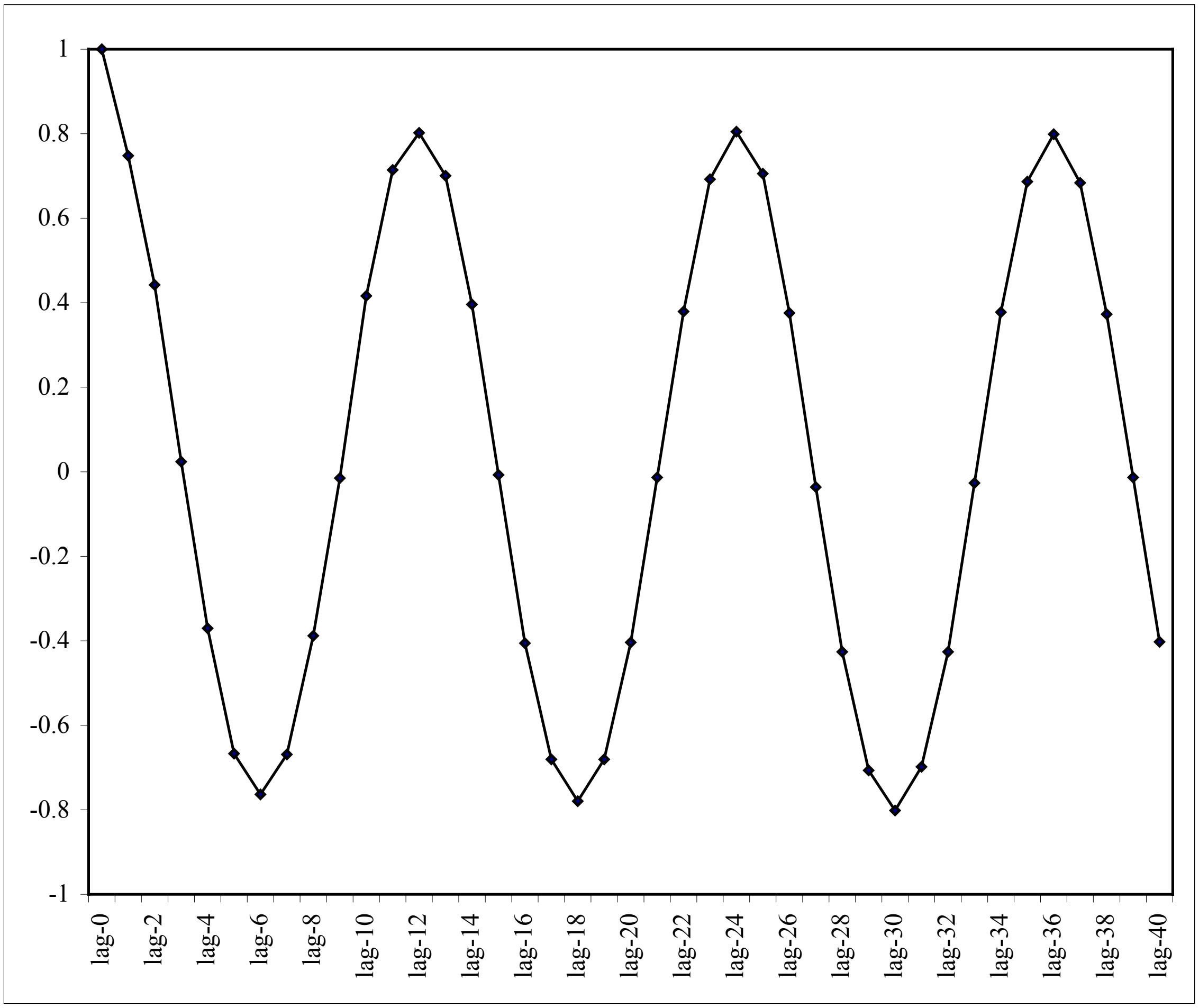

Fig.02-The autocorrelation function computed up to 40 lags.

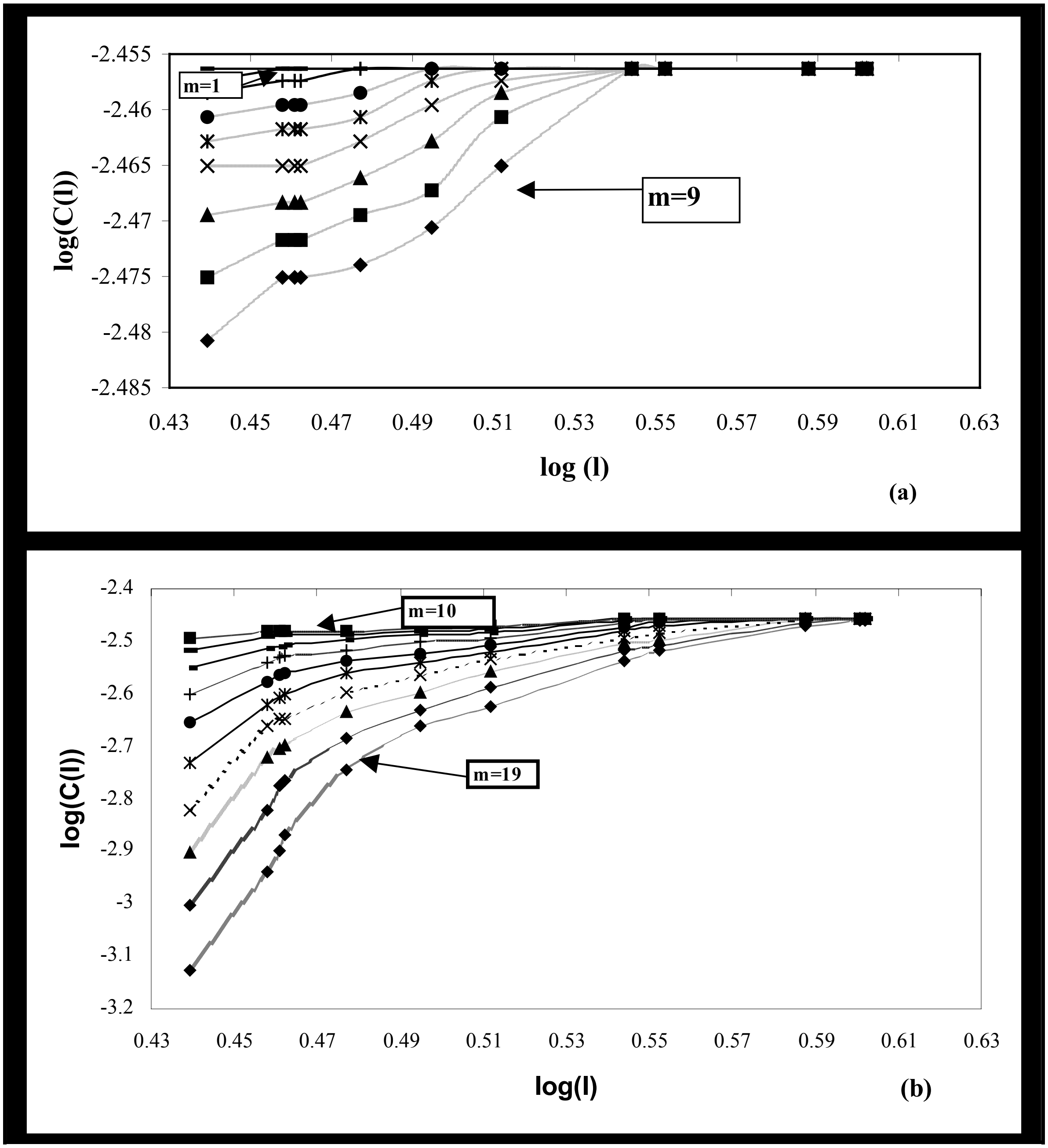

**Fig 03** (a) and (b) The *log* C(*l*) versus *log* (*l*) for the mean monthly total ozone time series over Arosa, Switzerland

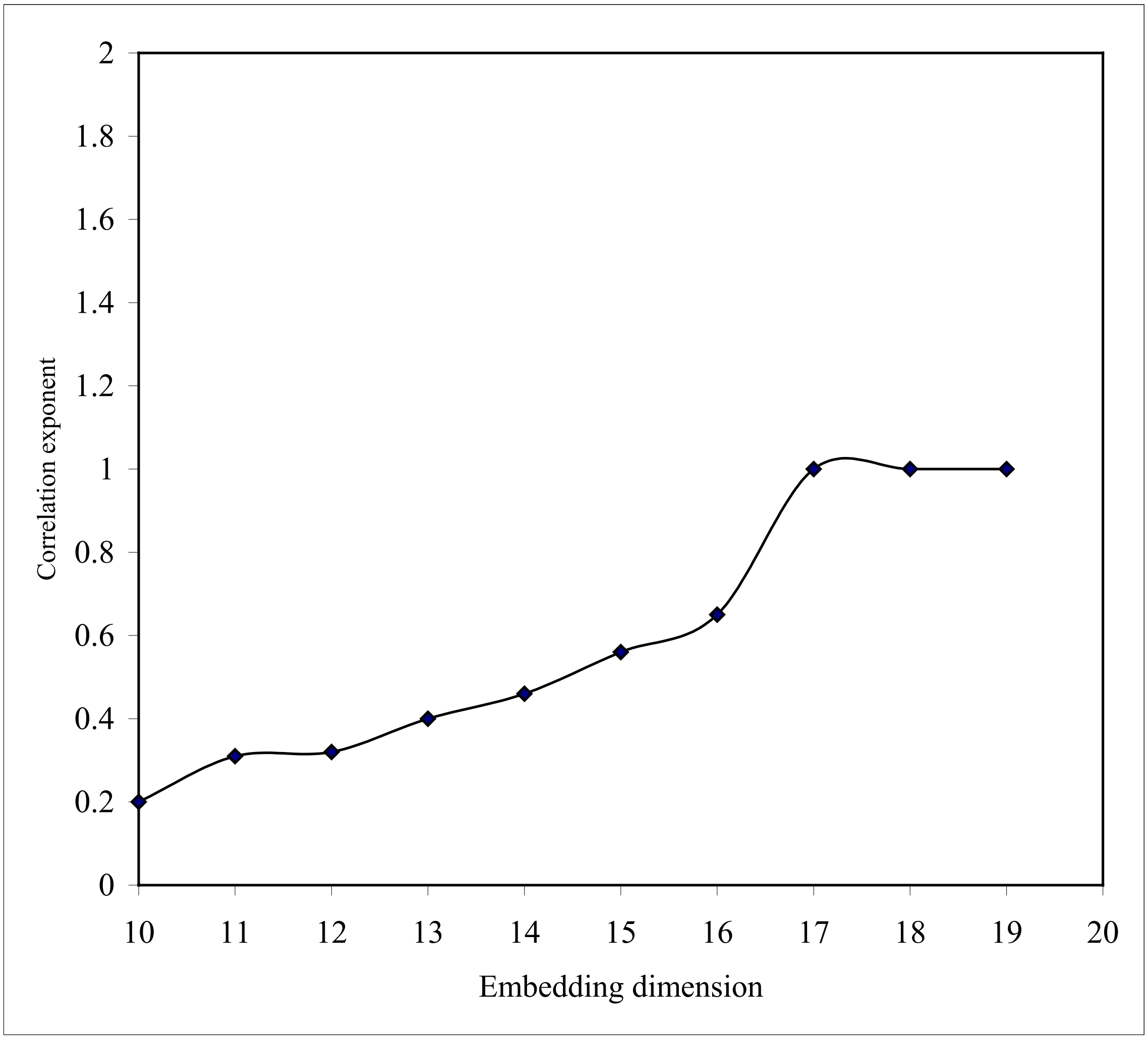

Fig.04- Relationship between correlation exponent and embedding dimension